\documentclass[13pt]{article}
\usepackage{graphics,epsfig,graphicx,amssymb,latexsym,amsmath}

\date{9 November 2003}

\newcommand {\fr}[2]{ \ensuremath{ \dfrac{#1}{#2} }}

\title{Benchmarking and Implementation of Probability-Based Simulations on Programmable Graphics Cards}
\author{S. Tomov, M. McGuigan, R. Bennett, G. Smith, J. Spiletic}

\begin{document}
\maketitle




\begin{abstract}
 The latest Graphics Processing Units (GPUs) are reported to reach up to 
 200 billion floating point 
 operations per second ($200$ Gflops \cite{advertisement}) and to have 
 price performance
 of $0.1$ cents per M flop. These facts raise great interest in
 the plausibility of extending the GPUs' use to non-graphics applications,
 in particular numerical simulations on structured grids (lattice). 
 We review previous work on using GPUs for non-graphics applications,
 implement probability-based simulations on the GPU, namely the
 Ising and percolation models, implement vector operation benchmarks
 for the GPU, and finally  compare the CPU's and GPU's performance. 
 A general conclusion from the results obtained is that 
 moving computations from the CPU to the GPU is feasible,
 yielding good time and price performance, for certain lattice computations.
 Preliminary results also show that it is feasible to use them in parallel.
\end{abstract}


\section{Introduction}
 There are several factors that motivated us to investigate the plausibility
 of using programmable GPUs for numerical simulations on 
 structured grids. These factors are the GPUs' 
 \begin{itemize}
   \item high flops count, 
   \item compatible price performance, and 
   \item better than the CPU rate of performance increase over time. 
 \end{itemize}
 To be specific, nVidia's NV30 graphics card is advertised
 to have a theoretical operation count of $200$ Gflops/s
 (see the NV30 preview \cite{advertisement} or Table \ref{TABLE1}, where
  we summarize the results of a GPU-CPU comparison in performing graphics
  related operations).
 Taking into account the price of the NV30 graphics card we get the low $0.1$ 
 cents per M flop. Also, the current development tendency shows a stable 
 doubling of the 
 GPU's performance every six months, compared to doubling the 
 CPU's performance for every $18$ months (a fact valid for
 the last $25$ years, known as {\it Moore's law}).

\begin{figure}[ht]
 \centerline{
    \includegraphics[width=5in]{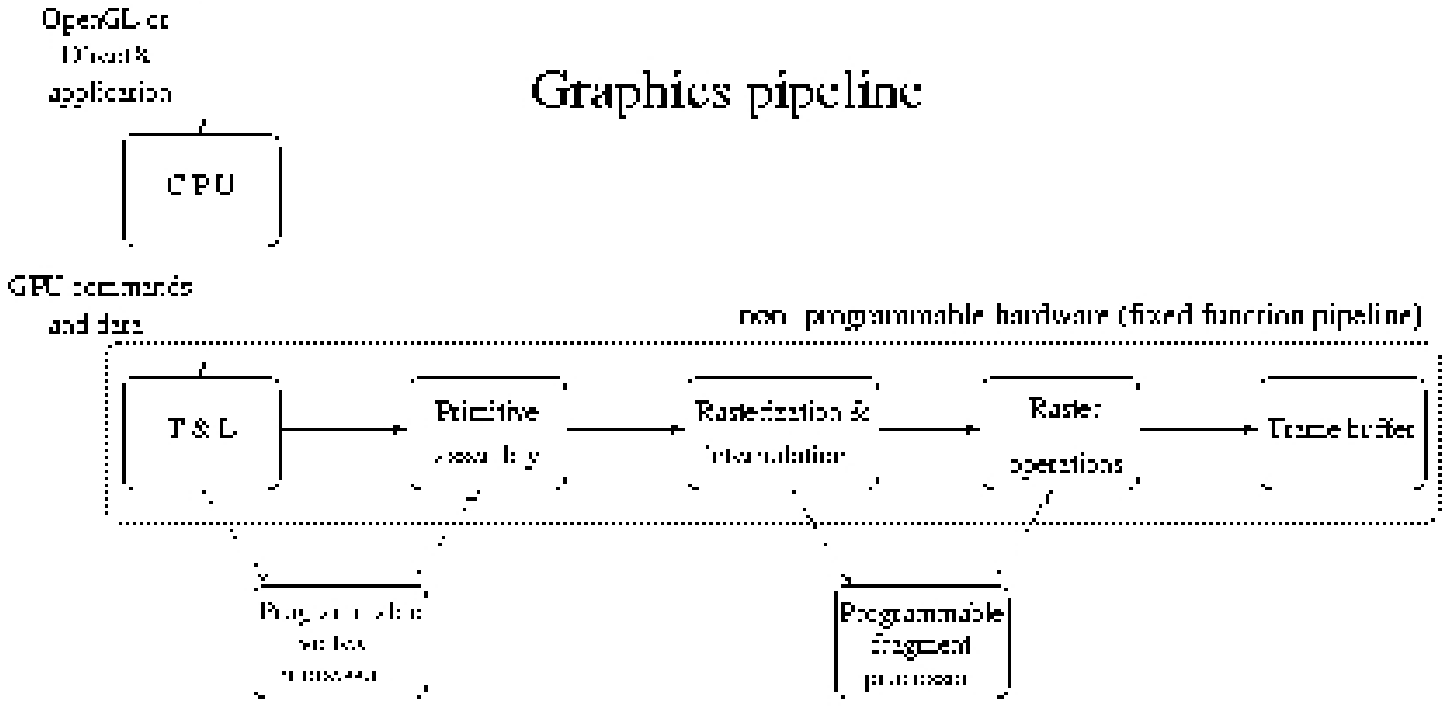}
 }

 \centerline{
    \includegraphics[width=5.1in]{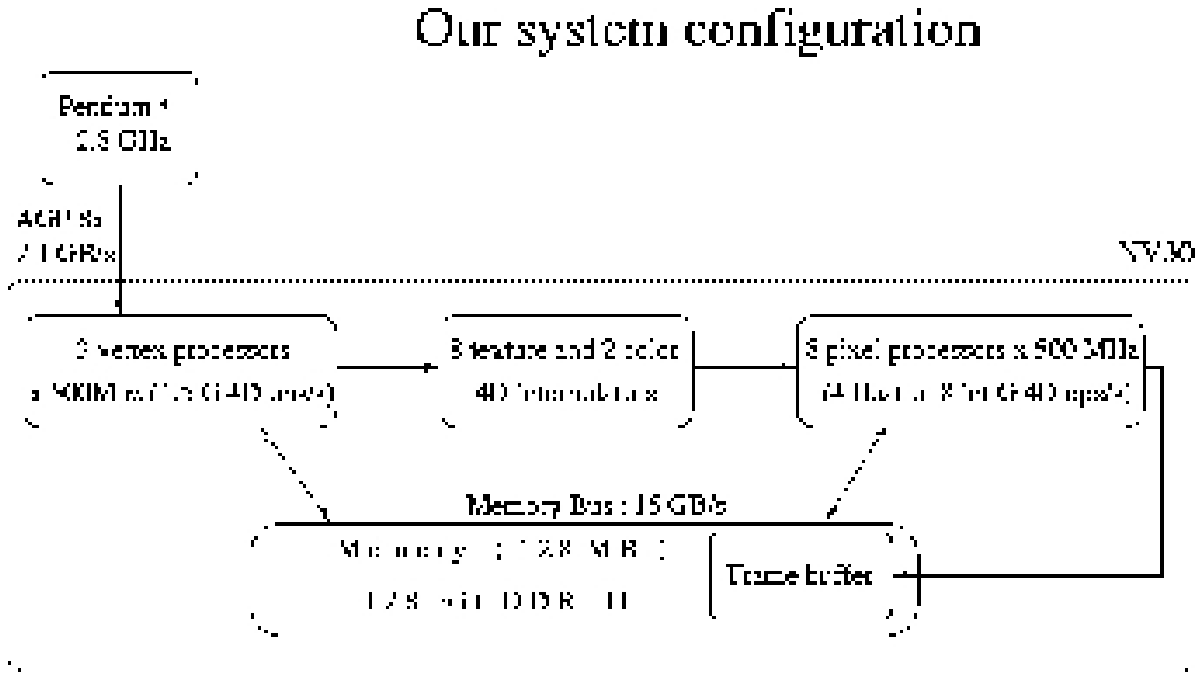}
            }
\caption{Up: Graphics pipeline. The dashed rectangle marks the fixed
         function graphics pipeline (with T\&L standing for transformation 
         and lighting). Additional hardware, programmable vertex and
         fragment processors, available in newer graphics cards provide 
         the developers with opportunity to change the fixed function 
         pipeline.
         Down: Our system configuration and CPU/GPU performance specifications
         (Pentium 4 CPU and NV30 GPU).}
\label{pipeline}
\end{figure}

Review of the literature on using GPUs for non-graphics applications
(subsection \ref{literature}) shows
that scientists report speedups of the GPU 
compared to the CPU for low precision operations. Several of the 
non-graphics applications that make use of the recently available 
$32$-bit floating point arithmetic report performance that is 
comparable to the CPU's performance. Thus, advertisements or scientific
papers that report speedups of one or above one order of magnitude in favor of
the GPU are restricted to low precision operations.
For example, the advertised $200$ Gflops/s in \cite{advertisement}
can not be achieved for $32$-bit floating point operations 
(Figure \ref{pipeline}, Right gives the 
 NV30 graphics card specifications).
Such advertisements and reports about ``extremely'' high performances,
combined with vagueness about the precision of the computations
motivated us to develop simple GPU benchmarking software. Benchmarking
the NV30 fragment processors with simple vector operations 
showed performance of $7$ Gflops/s which is $44\%$ of the 
theoretically possible maximal performance. 
This is the speed that we also achieved on
two probability-based applications that we implemented on the GPU.

\begin{table}[ht]
\begin{center}
\begin{tabular}{|c|c|c|} \hline
      & \multicolumn{2}{|c|}{Frames per second using}\\
\cline{2-3}
 \raisebox{1.0ex}[0pt]{Problem size} &  \raisebox{-0.5ex}[0pt]{OpenGL (GPU)} &
 \raisebox{-0.5ex}[0pt]{Mesa (CPU)}   \\ \hline 
 $11,540$  &  $189$             &    $8$           \\ \hline
 $47,636$  &  $52$              &    $1.71$        \\ \hline
 $193,556$ &  $13$              &    $0.44$        \\ \hline
 $780,308$ &  $3.28$            &    $0.12$        \\ \hline
\end{tabular}
\end{center}
\vspace{0.1in}
{\bf Table $1$.}~
   GPU {\it vs} CPU in rendering polygons.
   The GPU (Quadro2 Pro) is approximately $30$ times faster
   than the CPU (Pentium III, $1$ GHz) in rendering polygonal data 
   of various sizes.
\label{TABLE1}
\end{table}

Probability-based models, discussed in section \ref{probability},
have a wide area of applications. They are computationally intensive 
and lend themselves naturally to implementation on GPUs, as we show 
in the paper.

\subsection{Literature review}\label{literature}

The use of graphics hardware for non-graphics applications is 
becoming increasingly popular.
A few examples of such uses are 
matrix-matrix multiplication \cite{larsen},
visual simulations of boiling, convection, and reaction-diffusion
processes \cite{harris},
non-linear diffusion \cite{rumpf},
multigrid solver \cite{bolz,goodnight},
Lattice Boltzmann Method (LBM) \cite{li},
fast Fourier transform (FFT) \cite{moreland},  and
3D convolution \cite{hopf}. For a more complete list and more
information on general purpose computations that make use of the GPU see
the GPGPU's homepage:\\
{\tt http://www.gpgpu.org/}.

In all the cases the non-graphics computations
are expressed in terms of appropriate graphics operations.
These graphics operations are executed on the graphics card and
the results are used to interpret the results of the original non-graphics
computations. Up until recently the output of the graphics operations
was constrained to integers, which was a serious obstacle for
a meaningful use of many of the non-graphics algorithms developed
for graphics hardware. For example, M. Rumpf and R. Strzodka
\cite{rumpf} used 12-bit arithmetic (InfiniteReality2 graphics card)
for nonlinear diffusion problems, E. Larsen and D. McAllister
\cite{larsen} used 8-bit arithmetic (GeForce3 graphics card) 
for matrix-matrix multiplications, etc. The low accuracy of the graphics
cards computations encouraged research in complicated software techniques to 
increase the accuracy. See for example the range scaling and range separation
techniques developed in \cite{li}. Even when 16-bit floating
point precision became available (in GeForce4), it was not uniformly
provided throughout the graphics pipeline. This was the reason why
C. Thompson's et al. \cite{thompson} general-purpose vector operations 
framework was not able to retrieve the computational results 
without loss of precision.

Currently, graphics cards like the nVidia Quadro FX 1000, known also as
NV30, support 32-bit floating point operations throughout the
entire graphics pipeline. Features like programmable vertex 
and fragment shaders (see section \ref{programmable}) allow
software developers to easily modify the graphics pipeline,
which is used in most of the non-graphics applications.
The programmability also becomes easier since more alternatives to 
assembly programming become available. One example is the high level 
language Cg \cite{cgmanual}.

By porting non-graphics applications to the GPU, scientists try to achieve
computational speedup or upload some of the computations from the CPU by 
using the GPU as coprocessor. Currently significant speedups of GPU vs CPU
are reported if the GPU performs low precision computations. Depending on 
the configuration and the low precision used (often $8$-bit) scientists report
speedups in the range of $25-30$ times, and as high as $60$ times.
Advertisements that make claims such as that GeForce $4$ GPUs are
capable of $1.2$ trillion operations/s
or that a supercomputer made out of PlayStations is capable of $0.5$ trillion 
operations/sec \cite{ncsa} are based on certain types of 
low precision graphics operations.
Our experience shows that currently general purpose applications that use 
32-bit floating point arithmetic (on NV30) have a speed comparable to the 
CPU's (Pentium $4$, $2.8$ GHz). By comparable we mean that the GPU may be $2-6$
times faster for certain applications, but not ``outrageously'' (order of
magnitude) faster as seen in some low precision computations.
This was confirmed in our applications and
benchmarks, the multigrid solver in \cite{bolz}, the shaders' speeds
using $32$-bit floating point arithmetic from 
{\tt http://www.cgshaders.org}, etc. 

\subsection{Programmable graphics cards}\label{programmable}

Old graphics cards had a fixed function graphics pipeline,
a schematic view of which is given on Figure \ref{pipeline}, Up.
The operations in the dashed rectangle were configured on a very low
level and were practically impossible to change by software developers.
In August, $1999$ nVidia released the GeForce $256$ graphics card,
which allowed a certain degree of programmability of its pipeline. 
In February, 2001
nVidia released the GeForce3 GPU, which is considered to be the 
first fully programmable GPU. Here fully programmable means that 
developers were able to provide their own transformations and lighting 
(T \& L) operations (vertex shaders) to be performed on the vertices 
(by the Programmable vertex processor) 
and their own pixel shaders to determine the final pixels
color (executed on the Programmable fragment processor). 
Both the vertex and pixel shaders 
are small programs, which when enabled, replace the corresponding
fixed function pipeline. The programs get executed automatically for
every vertex/pixel and can change their attributes.
Originally the vertex and pixel shaders had to be written in assembly.
The constantly increasing functionality provided by the graphics 
cards allows for more complex shaders to be written. To simplify
the implementation of such shaders nVidia recently released 
a high level shader language, called Cg \cite{mark,cgmanual}. Cg stands for 
``C for graphics'' since it has syntax similar to C.

\section{Probability-based simulations}\label{probability}
In this section we briefly describe the models that we implemented on
the GPU, namely the Ising (Section \ref{ising}) and the percolation
(Section \ref{percolation}) models. Both methods are considered to
be of Monte Carlo type (Section \ref{monte_carlo}).

\subsection{Monte Carlo simulations}\label{monte_carlo}
Monte Carlo (MC) methods are used in the simulation of a variety of
phenomena in physics, finance, chemistry, etc. MC simulations are based on
probability statistics and use random numbers. The name derives from the 
famous Monte Carlo resort and is associated with roulette as a simple
way of generating random numbers.

A classical example of using MC methods is to compute the area of 
a circle. First the circle is circumscribed by a square, then random 
locations within the square are generated, and finally
\[
   \fr{\mbox{circle area}}{\mbox{square area}} ~ \approx ~ 
            \fr{\mbox{locations within the circle}}
               {\mbox{number of generated locations}},
\]
where the approximation greatly depends on the number of locations generated 
and the ``quality'' of the random locations generator. Our aim is to use
graphics cards for the fast generation of extremely large 
amounts of random numbers, and the model processing associated 
with the random numbers generated. To be specific, we would like 
to achieve performance close to the theoretically possible
maximal performance, which is $16$ Gflops for the fragment
processors of the NV30 graphics card.
  
A problem of general interest is the computation of expected values.
For example, assume we have a system that can be in any of its
states $S_i$, $i = 1,..., N$ with known probabilities $P(S_i)$. 
Also, assume a quantity of interest $F$ is computable for any of the
states. Then, the expected value for $F$, denoted by $E(F)$, is given by
\begin{equation}\label{expected}
    E(F) = \sum_{i=1}^{N} F(S_i) P(S_i).
\end{equation}
The Ising model, described in Section \ref{ising}, is a MC method for
computing such expected values. The difficulty in computing $E(F)$
is when $N$ becomes large. If we consider a $2$D system of particles,
say on a $1024 \times 1024$ lattice, and every particle is modeled to
have $k$ possible states, then $N$ is of order $k^{1024^2}$.

The accuracy of the Monte Carlo simulations, as mentioned above,
depends on the quality of the random number generator. Computer programs 
generate pseudo-random numbers. 
For the applications that we consider we used a linear congruential
type generator (LCG). The form is
\[
   R( n ) = \left(a * R(n-1) + b \right) \mbox{ mod } N.
\]
Fixing a starting value $R(0)$, called seed, uniquely determines the 
numbers generated.
LCGs are fast, easy to compute, and reasonably accurate. Furthermore, they 
lead to uniformly distributed random numbers  and are
the most frequently used. The LCGs are well understood and studied.
One has to be careful in the choice of the constants and the seed.
Undesirable patterns may occur in applications that consider
$n$-tuples of numbers generated by LCGs (see \cite{marsaglia}).

\begin{figure*}[ht]
  \centerline{
     \includegraphics[width=2.2in]{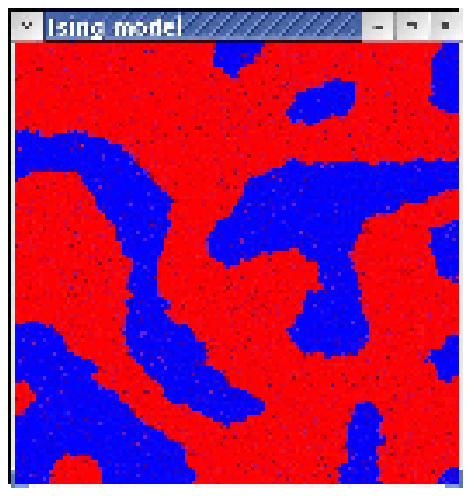}
     \hspace{0.5in}
     \includegraphics[width=2.2in]{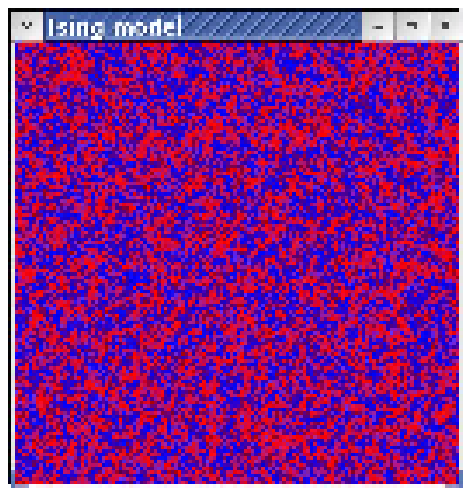}
             }
  \caption{Ising model visualization on lattices of size
           $512 \times 512$. The red/blue regions correspond to
           spins pointing up/down. Observe the spin clustering
           (related to energy minimization) for
           low input temperature (Left) and the randomness
           (related to entropy maximization) for higher 
           input temperatures (Right).}
\label{ising_visualization}
\end{figure*}

\subsection{Ising model}\label{ising}
The Ising model, a simplified model for magnets, was introduced by Wilhelm 
Lenz in 1920 and further developed later by his student Ernst Ising.
The model is on a lattice, in our case two dimensional. A ``spin'', pointing
up or down, is associated with every cell of the lattice. The spin 
corresponds to the orientation of electrons in the magnet's atoms. There are 
two opposing physics principles that are incorporated in the Ising
model: (1) minimization of the system's energy, achieved by spins pointing
in one direction, and (2) entropy maximization, or randomness, achieved 
by spins pointing in different directions. The Ising model uses
temperature to couple these opposing principles (see an illustration
on Figure \ref{ising_visualization}).
There are many variations of the Ising model and its implementation 
(see \cite{creutz} and the literature cited there).
The computational model that we used is described as follows.

We want to be able to compute various expected values (equation 
\ref{expected}), such as expected magnetization and expected energy.
To compute expected magnetization, $F(S_i)$ from equation (\ref{expected})
is the magnetization 
of state $S_i$, defined as the number 
of spins pointing up minus the number of spins pointing down.
To compute expected energy, $F(S_i)$ is the energy 
of state $S_i$
\[
   \mbox{E}(S_i) = - \sum_{<j,k>} S_i(j) S_i(k),
\]
where the sum is over all lattice edges with $<j,k>$ being the edge connecting
nearest neighbor sites $j$ and $k$, $S_i(j)$ is the spin for site
$j$ of state $S_i$ with values $1$, for spin pointing for example up, 
or $-1$, for spin pointing down. 

The idea is not to compute the quantity described in equation 
(\ref{expected}) exactly, since many of the states are of very low 
probability, but to evolve the system into ``more probable'' states
and get the expected value as the average of several such ``more probable''
states. The user inputs an absolute temperature of interest $T$ in Kelvin, 
and probability 
$p \in [0,1]$ for spins pointing up. Using this probability we generate 
a random initial state with spins pointing up with probability $p$. 
The procedure for evolving from this initial
state into ``more probable'' states is described in the following
paragraph. The theoretical justification of methods dealing with 
similar sequences evolving from state to state, 
based on certain probability decisions, is related
to the so-called {\it Markov chains} \cite{markov}.

The lattice is colored in a checkerboard manner. We define a sweep as
a pass through all white or all black sites. This is done so that the
order in which the sites are processed does not matter. We start consecutive
black and white sweeps. At every site we make a decision whether to 
flip its spin based on the procedure
\begin{enumerate}
   \item Denote the present state as $S$, and the state with flipped
          spin at the current site as $S'$.
   \item Compute $\triangle \mbox{E} \equiv \mbox{E}(S') - \mbox{E}(S)$.
   \item If $\triangle \mbox{E} < 0$ accept $S'$ as the new state.
   \item If $\triangle \mbox{E} \geq 0$, generate a random number 
         $R \in [0, 1]$, 
         and accept $S'$ as the new state if 
            \[
                 R \leq \fr{P(S')}{P(S)} = 
                  e^{\small{ -\triangle \mbox{E}/(k T)}},
            \]
         otherwise the state remains $S$. In the last formula
         the probability $P(S)$ for a state $S$ is given by
         the Boltzmann probability distribution function
            \[
                P(S) = \fr{e^{\small{ -\mbox{E}(S)/(k T)}}}{
                 \sum\limits_{i=1}^N~ e^{\small{ -\mbox{E}(S_i)/(k T)}}} ~,
            \]
         where $k$ is a constant, known as the Boltzmann's constant.
\end{enumerate}  

A standard Monte Carlo implementation of the Ising model would involve 
a random traversing through different sites of the lattice, and flipping 
or not the spin depending on the outcome of the above procedure. The
implementation that we describe is influenced by one of M. Creutz's 
\cite{creutz} simulations of the Ising model. He uses the 
checkerboard sweeps in a fully deterministic spin flipping dynamic,
which does not require generation of high quality random numbers and
yields one order of magnitude faster execution than conventional
Monte Carlo implementations. For proof of concept in benchmarking the GPU
we use the algorithm described, which involves a richer variety of
mathematical operations. 
The checkerboard sweeps are crucial for the efficient use of the GPU,
since the GPU computations are performed on passes over the entire
lattice. Simultaneous update on all the sites would not be a simulation 
of the Ising model, as shown in \cite{vichniac}.

\subsection{Percolation model}\label{percolation}
The percolation model is a model for studying disordered media.
It was first studied by Broadbent and Hemmercley\cite{percolation} 
in $1957$. Since then percolation type models 
have been used in the study of various phenomena, such as the spread of
diseases, flow in porous media, forest fire propagation, phase
transitions, clustering, etc. 
The general scenario is when a medium is modeled as an interconnected
set of vertices (sites). Values modeling the media and the 
phenomena of interest
are associated with the sites and the connections between them.
Usually these values are modeling disordered media with a certain probability 
distribution, 
and are obtained from random number 
generators. A point (or points) of ``invasion'', such as the starting
point of fire, disease, etc, is given, and based on
the values in the sites and the connections, a probabilistic action 
for the invasion is taken. Of particular interest is to determine:
\begin{enumerate}
  \item Media modeling threshold after which there exists a 
        spanning (often called percolating) cluster. This is an interconnected
        set of invaded sites that spans from one end of the medium to another.
  \item Relations between different media models and time to reach 
        steady state invasion or percolation cluster. 
\end{enumerate}

Using the basics of the percolation, as described above,
one can derive methods of substantial complexity. For example,
R. Saskin et al. \cite{sasik} presented a novel approach for the clustering
of gene expression patterns using percolation: gene
expressions are modeled as (1) sites containing $m$ measurements each,
and (2) connections between all pairs of sites. The connections
have weights which represent the similarity between the gene expressions.
Based on the mutual connectivity of the gene expressions, they come up
with a probabilistic percolation model to cluster the data.

\begin{figure}[ht]
 \centerline{
    \includegraphics[width=2.2in]{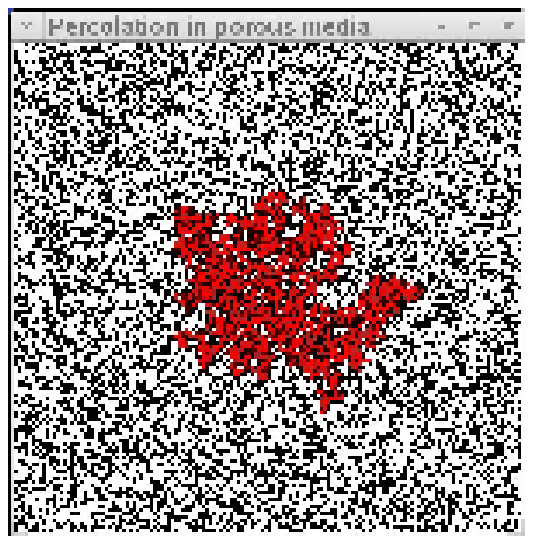}
            }
\caption{Percolation model visualization. Seen is a snapshot of 
         fluid spreading in 2D porous media (modeled on a lattice of size
         $512 \times 512$, porosity $0.6$).}
\label{percolation_visualization}
\end{figure}

Here,
for proof of concept in benchmarking the GPU, we 
implemented a simple percolation model with application to 
diffusion through porous media (see Figure \ref{percolation_visualization}). 
The porous media is modeled
on a two dimensional lattice. First, we specify
porosity as a probability $P$ of the site being a pore.
Then we go through all the sites and at each site (1) generate a 
random number  $R \in [0,1]$, and (2) consider the site ``pore'' if
$R \leq P$, or solid otherwise. 
Finally, we consider an invasion point, initialize a cluster to be
the invasion point, and start the process of spanning the initial 
cluster through the porous space.

\section{Implementation details}\label{implementation}

The implementations of the models described are Cg fragment programs, 
which are being invoked (and executed on the GPU) by OpenGL application 
running on the CPU. To do the OpenGL - fragment programs binding we used 
the OpenGL Cg run-time functions described in the Cg toolkit user's manual 
\cite{cgmanual}. The programming model is described in the following
paragraph.

The OpenGL application repeatedly executes the iterative procedure:
\begin{itemize}
\item Prepare the input for the GPU pipeline;
\item If the GPU executes a non-fixed function pipeline specify 
      the vertex or fragment programs to be used;
\item If needed, read back the graphics pipeline output
      (or part of it).
\end{itemize}
This programming model is standard. It is very often related to the 
so-called {\sl Dynamic texturing}, where on the first step one creates a
texture T, then a GPU fragment program uses T in rendering an image in
an off-screen buffer (called p-buffer),
and finally T is updated from the resulting image.

To be more specific, we have used GL\_TEXTURE\_RECTANGLE\_NV
textures, an extension to the 2D texture, which allows 
the texture dimensions not to be a power of 2. To initialize such
textures from the main memory we use 
glTexSubImage2D(GL\_TEXTURE\_RECTANGLE\_NV, ... ). To
replace (copy) 
rectangular image results from the p-buffer to 
GL\_TEXTURE\_RECTANGLE\_NV textures we use
glCopyTexSubImage2D(GL\_TEXTURE\_RECTANGLE\_NV, ...),
which is entirely done on the graphics card.
We used the floating point p-buffer that is distributed with
the Cg toolkit which is available on the Cg homepage \cite{cg_homepage}.
For our fragment programs we used fp30 profile with options set by
the cgGLSetOptimalOptions function.

The times in the benchmarking programs were measured with the
gettimeofday function. Since the OpenGL calls are asynchronously
executed we used glFinish() to enforce OpenGL requests completion
before measuring their final execution time. The time measures were used 
to determine the GPU's performance.

\section{Benchmarking the GPU}\label{benchmarking}

We wrote simple fragment programs to benchmark the performance of
simple floating point vector operations on the GPU. 
The results were compared 
with those on the CPU.
The specifications of the system on which we applied the benchmark
programs are given in Figure \ref{pipeline}, Down.
We perform vector operations, where the vectors are represented by
2D textures. Here we included computational results for textures
of sizes $256\times256$ and $512\times512$ (with corresponding vector sizes equal
to texture sizes times $4$). 
\begin{table}[ht]
\begin{center}
\begin{tabular}{|c|c|c|} \hline
      & \multicolumn{2}{|c|}{Lattice size (not necessary power of $2$)}\\
\cline{2-3}
\raisebox{1.0ex}[0pt]{operation} & \raisebox{-0.5ex}[0pt]{~~~$256 \times 256$~} & 
                                   \raisebox{-0.5ex}[0pt]{~$512 \times 512$~}\\ \hline 
   =             & 0.00063 & 0.0024  \\ \hline
  + , - , *, /   & 0.00073 & 0.0027  \\ \hline
  $cos$, $sin$   & 0.00089 & 0.0034  \\ \hline
  $log$, $exp$   & 0.00109 & 0.0039  \\ \hline
\end{tabular}
\end{center}
\vspace{0.1in}
{\bf Table $2$.}~
   Time in seconds for different floating point vector operations on the 
   GPU (NV30). The vector sizes are lattice size times $4$.
\label{GPUflops_performance}
\end{table}

Table \ref{GPUflops_performance} gives the GPU 
performance for different floating point vector operations. 
Table \ref{CPUflops_performance} is the corresponding CPU 
performance.

\begin{table}[ht]
\begin{center}
\begin{tabular}{|c|c|c|} \hline
      & \multicolumn{2}{|c|}{Lattice size}\\
\cline{2-3}
\raisebox{1.0ex}[0pt]{operation}& \raisebox{-0.5ex}[0pt]{~~~$256 \times 256$~} & 
                                  \raisebox{-0.5ex}[0pt]{~$512 \times 512$~}\\ \hline 
+ , - , *, /   & 0.0011 & 0.0046  \\ \hline
$cos$ , $sin$  & 0.0540 & 0.0650  \\ \hline
$log$, $exp$   & 0.0609 & 0.1100  \\ \hline
\end{tabular}
\end{center}
\vspace{0.1in}
{\bf Table $3$.}~
   Time in seconds for different floating point vector operations on the 
   CPU (2.8 GHz Pentium 4). The vector sizes are lattice size times $4$.
\label{CPUflops_performance}
\end{table}

The performance is measured as explained in section \ref{implementation}.
We performed our computations and measures on Linux Red Hat 9 operating 
system. Problems related to floating point p-buffers were observed 
with nVidia driver display 4191 from December, 2002.
These were fixed in the drivers that followed. 
The computations in Tables \ref{GPUflops_performance} and 
\ref{CPUflops_performance}, and sections \ref{results} and 
\ref{extensions} are done using Linux driver display 4363 from April, 2003.
Recently we installed the newest driver, version 4496 from July, 2003, 
and observed a speedup of approximately 2 times compared to the 
older driver.

The operations measured are in the following forms
$a=c$ (for the $=$ operation), $a+=c$ (for the basic arithmetic operations), 
and $a=sin(a)$ (for the $sin$, $cos$, $log$, and $exp$ operations),
where $a$ is a vector represented with a 2D texture and 
$c$ is a constant. The $a=c$ operation (Table \ref{GPUflops_performance})
is present in all of the operations considered. 
Its execution time includes overheads related to
the graphics pipeline and time for transferring data 
(reading and writing to the local GPU memory of four 32-bit 
floating point values for each lattice vertex). Having the time
for this operation allows as to exclude it from the other operations
and get a better idea about their performance. For example,
adding a basic arithmetic operation to a fragment program will increase
the execution time on a lattice of size $512 \times 512$ with $0.0003$ 
seconds. Thus the performance would be $512^2*4/0.0003 \approx 3.5$
Gflops/s. This is the performance to be expected for longer fragment
programs, where the overhead time would get significantly small. Indeed,
in section \ref{results} we achieve this performance for the Ising
model, which has a fragment program of $109$ assembly instructions.
The overhead time is significantly large compared to the other operations,
which shows that an approach of making a library of different basic vector 
operations on the GPU would not be efficient. For example the
basic arithmetic operations on a $512 \times 512$ lattice would
yield a performance of approximately $0.39$ Gflops/s instead
of the much higher $3.5$ Gflops/s.

\begin{table*}[ht]
\begin{center}
\begin{tabular}{|c|c|c|c|c|c|} \hline
      & \multicolumn{5}{|c|}{Lattice size (not necessary power of 2)}\\
\cline{2-6}
              &  $64\times64$ & $128\times128$ & $256\times256$ & $512\times512$ & $1024\times1024$ \\ \hline
GPU sec/frame & 0.0006 & 0.0023  & 0.0081  & 0.033   & 0.14      \\ \hline
CPU sec/frame & 0.0008 & 0.0020  & 0.0069  & 0.026   & 0.10      \\ \hline
\end{tabular}
\end{center}
\vspace{0.1in}
{\bf Table $4$.}~
   GPU (NV30) and CPU (2.8 GHz Pentium 4) performances in 
   processing a single step of the Ising model.
\label{ising_performance}
\end{table*}

Overheads related to data traffic are also observed on the CPU, but they
are more difficult to measure. For example the basic operations would 
take less time to compute than $a=c$ (we use full compiler optimization).
This suggests that the time of the basic operations is a measure
for these overheads. Note that the $cos$, $sin$, $log$, and $exp$ are 
significantly faster on the GPU. Part of the reason is that they are 
accelerated on the GPU by reducing the precision of their computation.

 The results from the other benchmark problems that we developed,
namely the Ising and percolation implementations, are given
in section \ref{results}. Measures for the CPU-GPU communication speed
are given in section \ref{extensions}, Table \ref{communication_performance}.
The communications measured use the AGP 8x port, which has a theoretical
bandwidth of $2.1$ GB/s on the NV30 
(Figure \ref{pipeline}, Down).

\section{Performance results and analysis}\label{results}

We tested our Ising model implementation with results from Michael Creutz's 
\cite{creutz} implementation. As expected, for lower input temperatures
the minimization of the system's energy physical principle prevails,
which is expressed by clustering of spins pointing in one direction
(see Figure \ref{ising_visualization}, Left). For higher temperatures
the entropy maximization physical principle prevails,
which is expressed by randomness in the spin orientations
(see Figure \ref{ising_visualization}, Right).

We tested our percolation model implementation by experimentally 
confirming the well known fact that the percolation
threshold, or in our case the porosity threshold for fluid flow, in 2D is 
$0.592746$ (see \cite{jan}). Figure \ref{percolation_visualization} gives
a snapshot of fluid spreading in 2D porous media.

Comparison between the GPU and CPU performances on the Ising model
is given in Table \ref{ising_performance}. The CPU code is compiled
with full optimization.
The results show that the GPU and the CPU have comparable speed for this 
application, with the CPU being slightly faster.
With the fast developments in the graphics hardware, it is
no surprise that optimization opportunities may be missed at the 
software level. Indeed, as mentioned in section \ref{benchmarking}, 
updating the display driver increased the GPU's performance approximately
by a factor of $2$. The Ising fragment program gets compiled using
the Cg compiler (cgc -profile fp30 ising.cg) to $109$ instructions.
Thus the achieved GPU performance, for example on a lattice of size 
$256 \times 256$, is $256^2*109*4/0.0081 \approx 3.5$ Gflops/s using 
the old display driver and
$7$ Gflops/s using the latest display driver, which is
correspondingly $22\%$ and $44\%$ of the theoretically
maximal floating point performance of the NV30 GPU
(theoretically the NV30 can achieve $16$ Gflops/s, 
see Figure \ref{pipeline}, Down).

One performance drawback in our Ising model implementation
is due to the fact that currently GPUs, and in particular the NV30,
do not support branching in the fragment programs (see 
\cite{cgmanual}, Appendix C, Step 9). This means that a 
conditional {\tt if/else} statement would get executed for the time
as if all the statements were executed, regardless of the 
condition. In our case we implemented the checkerboard execution
pattern with conditional statement in the fragment program.
Modeling the checkerboard pattern with 2 lattices (one for
black and one for white squares) and properly modifying the code 
could replace the {\tt if/else} condition in the fragment program
and thus increase the speed by a factor of $2$.
Another important consideration related to the GPU's performance
is that the operations should be organized in terms of 4D vector
operations. For a full list of considerations for achieving high 
performance see \cite{cgmanual}, Appendix C.

\section{Extensions and future work}\label{extensions}
 
Our implementation of the Ising and the percolation models 
were designed mostly to prove a concept and to be used in
benchmarking the GPUs floating point performance. The
implementation goal skipped some optimization issues that
are important in developing GPU software. Currently we are
looking closely at the assembly code generated by Cg and are
working on its optimization.  Another direction is to increase the number
of applications for the GPUs. We are working on 
Quantum Chromodynamic (QCD) applications and fluid
flow simulations. Such applications usually lead to 
large scale computations which are impossible to 
perform on a single GPU (or CPU) due to memory constraints.
Therefore, a main direction of our research is on the parallel
use of programmable GPUs. Table \ref{communication_performance}
gives the different communication rates between the CPU and the 
GPU for lattices of different sizes. Although the results are
far from the theoretical maximum, AGP $8$x graphics bus with bandwidth 
$2.1$ GB/s, the CPU-GPU communication speed will not be a bottleneck in 
a commodity-based cluster. 

\begin{table*}[ht]
\begin{center}
\begin{tabular}{|c|c|c|c|c|c|} \hline
          & \multicolumn{4}{|c|}{Lattice size (not necessary power of 2)} & 
                                                         $\approx$ speed \\
\cline{2-5}
\raisebox{1.0ex}[0pt]{Operation} & \raisebox{-0.5ex}[0pt]{~~$64 \times 64$~} &
                                   \raisebox{-0.5ex}[0pt]{$128 \times 128$} &
                                   \raisebox{-0.5ex}[0pt]{$256 \times 256$} &
                                   \raisebox{-0.5ex}[0pt]{$512 \times 512$} &
(MB/s) \\ \hline
Read bdr   & 0.00016 & 0.0002 & 0.0006 & 0.0024 &~14 \\ \hline
Read all   & 0.00040 & 0.0015 & 0.0062 & 0.0250 &167 \\ \hline
Write bdr  & 0.00022 & 0.0003 & 0.0007 & 0.0024 &~14 \\ \hline
Write all  & 0.00020 & 0.0008 & 0.0032 & 0.0120 &350 \\ \hline
\end{tabular}
\end{center}
\vspace{0.1in}
{\bf Table $5$.}~
   Communication rates between the CPU and the GPU for lattices of 
   different sizes. The reading is done using the glReadPixels function,
   writing the boundary is done using the glDrawPixels function, and writing 
   the whole domain is done using the glTexSubImage2D function.
\label{communication_performance}
\end{table*}

\section{Conclusions}\label{conclusions}
Our analysis and benchmarking were mainly concentrated on
the NV30 fragment processors' floating point operations performance. 
The results show that it is feasible to use GPUs for numerical simulations. 
We demonstrated this by banchmarking the GPU's performance on 
simple vector operations and by implementing two probability-based 
simulations, namely the Ising and the percolation models. For these two 
applications, the applications cited in the literature overview, our
vector operations benchmarks, and various nVidia 
distributed shaders we observed that the fragment processors' (nVidia NV30) 
floating point performance is comparable to the performance of a Pentium 4 
processor running at 2.8 GHz. By comparable we mean that the GPU may be $2-6$
times faster for certain applications. For example, we achieved $44\%$ of the 
theoretically possible maximal performance. Such performance made
the GPU's implementation run twice faster than the CPU's. A modification
of the algorithm that removes the conditional {\tt if/else} statements
related to the checkerboard type traversal of the 2D lattices would
contribute an additional speedup of at least two times.
These results make the use of the GPU as a coprocessor appealing. 
Also, GPUs tend to have a higher rate of performance increase 
over time than the CPUs, thus making the study of
non-graphics applications on the GPU valuable research for the future.
Speedups of approximately one 
order of magnitude in favor of the GPU are observed only in certain 
applications involving low precision computations. A reason for 
this difference is the larger traffic involved in higher precision 
computations which traffic makes the GPUs' local memory bandwidth
a computational bottleneck. 
%
%
Finally, we note that the acceleration graphics ports 
provide enough bandwidth for the CPU-GPU communications to
make the use of parallel GPUs computations feasible.

\section*{Acknowledgments}
We would like to thank Beverly Tomov from Cold Spring Harbor Laboratory, 
NY, for her attentive editing and remarks. 
We thank Michael Creutz from Brookhaven National Laboratory, NY, for the 
discussions with him, his suggestions, and advice about the
models implemented.


\end{document}